\documentclass[aps,prx,twocolumn,showpacs,superscriptaddress,floatfix]{revtex4-2}

\usepackage{graphicx,graphics}
\usepackage{dcolumn}

\usepackage{amsmath,amssymb,amsfonts}
\usepackage{latexsym,verbatim}
\usepackage{bm}
\usepackage{lipsum} 
\usepackage{dsfont}
\usepackage{cancel}
\usepackage{breqn}
\usepackage{tikz}
\usepackage{bbm} 

\makeatletter
\let\cat@comma@active\@empty
\makeatother
\usepackage{wasysym}

\usepackage[toc,page]{appendix}
\usepackage{color}
\usepackage{diagbox}
\usepackage{ulem}
\usepackage{tikz}
\usepackage[breaklinks=true,colorlinks,citecolor=blue,linkcolor=blue,urlcolor=blue]{hyperref}
\usepackage{tabularx}
\usepackage{array, multirow, bigdelim, makecell, booktabs} 
\graphicspath{fig}

\begin{document}
\title{Understanding Symmetry Breaking in Twisted Bilayer Graphene \\ from Cluster Constraints}

\author{Nikita~Astrakhantsev}
\email[]{nikita.astrakhantsev@physik.uzh.ch}
\affiliation{Department of Physics, University of Zurich, Winterthurerstrasse 190, CH-8057 Z\"{u}rich, Switzerland}

\author{Glenn~Wagner}
\affiliation{Department of Physics, University of Zurich, Winterthurerstrasse 190, CH-8057 Z\"{u}rich, Switzerland}

\author{Tom~Westerhout}
\affiliation{Institute for Molecules and Materials, Radboud University, Heyendaalseweg 135, 6525AJ Nijmegen, The Netherlands}

\author{Titus~Neupert}
\affiliation{Department of Physics, University of Zurich, Winterthurerstrasse 190, CH-8057 Z\"{u}rich, Switzerland}

\author{Mark~H.\,Fischer}
\affiliation{Department of Physics, University of Zurich, Winterthurerstrasse 190, CH-8057 Z\"{u}rich, Switzerland}

\begin{abstract}
Twisted bilayer graphene is an exciting platform for exploring correlated quantum phases, extremely tunable with respect to both the single-particle bands and the interaction profile of electrons. Here, we investigate the phase diagram of twisted bilayer graphene as described by an extended Hubbard model on the honeycomb lattice with two fermionic orbitals (valleys) per site. Besides the special extended {\it cluster interaction} $Q$, we incorporate the effect of gating through an onsite Hubbard-interaction $U$. Within Quantum Monte Carlo (QMC), we find valence-bond-solid, Néel-valley antiferromagnetic or charge-density wave phases. Further, we elucidate the competition of these phases by noticing that the cluster interaction induces an exotic constraint on the Hilbert space, which we dub {\it the cluster rule}, in analogy to the famous pyrochlore spin-ice rule. Formulating the perturbative Hamiltonian by projecting into the cluster-rule manifold, we perform exact diagonalization and construct the fixed-point states of the observed phases. Finally, we compute the local electron density patterns as signatures distinguishing these phases, which could be observed with scanning tunneling microscopy. Our work capitalizes on the notion of cluster constraints in the extended Hubbard model of twisted bilayer graphene, and suggests a scheme towards realization of several symmetry-breaking insulating phases in a twisted-bilayer graphene sheet.
\end{abstract}

\maketitle
\section{Introduction}
\label{sec:intro}

Twisted bilayer graphene (TBlG) has emerged as a versatile platform for studying competing phases in a system with strong correlations. Experiments have reported correlated insulating phases~\cite{Park_2021,Cao2018,Yankowitz_2019,Cao_2021,Sharpe_2019,Serlin900,Lu2019,Stepanov_2020,Wu_2021,Zondiner_2020}, in other words insulating phases where band theory predicts a metal, as well as nodal (unconventional) superconductivity~\cite{Cao2018,Cao2018b,Yankowitz_2019,Lu2019,Oh2021}. The relationship between the insulators and the superconducting phases is still unclear: either the insulator can be viewed as the parent state, which upon doping becomes superconducting, or the insulator and superconductor are competing phases, with different underlying mechanisms.

Experiments have addressed this question by studying the dependence of the phase diagram of TBlG on the electronic screening. Increasing the screening decreases the range and strength of the Coulomb interaction. In Ref.~\onlinecite{Li2021}, the screening was tuned by varying the electron density in a metal close to the TBlG, while in Refs.\,\onlinecite{Saito2020,Stepanov_2020} distinct devices with different gate distances were investigated. Strikingly, the experiments observed that the insulating phases weaken or disappear, while superconductivity survives even as the screening is increased. In our work, we focus on understanding the effect of the modified screening on the correlated insulators. 

TBlG sits at the intersection of two paradigmatic models of strongly correlated phases. On the one hand, the flat Chern bands are reminiscent of the Landau levels of the quantum Hall effect, an analogy that can be made precise in an idealized model of TBlG~\cite{Tarnopolsky}. In this language, the correlated insulators are generalized quantum Hall ferromagnets that may exhibit intervalley coherence (IVC)~\cite{Bultinck,IKS_PRX,Lian,Xie}. On the other hand, the proximity of correlated insulating and unconventional superconducting phases as well as linear-in-$T$ resistivity~\cite{Jaoui2022,Polshyn2019} suggests a connection to the phenomenology of the Hubbard model used to model cuprate superconductors. In that language, the correlated insulators are valence bond solid (VBS) or quantum valley Hall (QVH) phases~\cite{qmc_paper,Xu,Breio}. Aiming at studying a strong coupling theory with local constraints, in this work we consider the Hubbard model of TBlG with realistic extended interactions. In the phase diagram of this model, we identify the emergent strongly-correlated phases and discuss how to tune the interactions experimentally across the phase diagram. We emphasize the similarity of these extended interactions' ground states to the iconic spin-ice manifold in the pyrochlore lattice~\cite{Hermele_2004} and develop an intuitive perturbation theory for hopping terms. Finally, we provide possible STM signatures of these correlated phases.

The special extended interactions emerge due to the characteristic `fidget spinner' form~\cite{Kang,PhysRevX.8.031087,Po} of the maximally localized Wannier functions. This leads to significant overlaps between Wannier orbitals on neighbouring sites implying that longer-range Hubbard interactions need to be included. The Coulomb interaction leads to the {\it cluster-interaction} $Q$, where the onsite, nearest-neighbour, next-nearest-neighbour and next-to-next-nearest-neighbour interactions satisfy the ratios 3:2:1:1. This interaction can be written as a sum of perfect squares fixing the total charge per hexagon (cluster), thus the name. Besides, to mimic the screening we introduce an on-site interaction $U$ that we tune separately~\footnote{The choice to consider only the change in the on-site interaction is motivated by the types of interaction terms that are simple to simulate within the quantum Monte-Carlo.}.

\begin{figure*}[t!]
    \includegraphics[width=\textwidth]{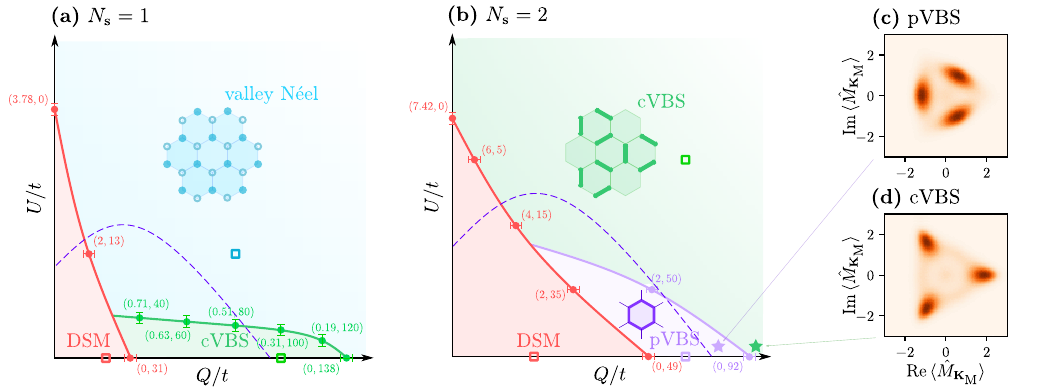}

    \caption{Phase diagram in the $(Q / t, U / t)$ plane. The fixed-point states are sketched in the respective regions of the phase diagram. The $(U, Q)$ pairs mark the phase boundary obtained by extrapolating the QMC data to the thermodynamic limit. The error bars indicate the direction of a scan in the plane. The uncertainty is $\Delta Q / t = 1.0$ for horizontal and $\Delta U / t = 0.01$ for vertical scans. Some transition points in these phase diagrams are already known from the literature: the $U / t = 0$ axis at $N_{\text{s}} = 2$ was studied in Ref.\,\cite{qmc_paper}, the $Q / t = 0$ axes at $N_{\text{s}} = 1$ and $2$ were studied in Refs.\,\cite{Assaad_2013,assaad_lectures}. The squares identify the points of the phase diagram where we later show infinite-volume extrapolations of susceptibilities within QMC. The dashed violet lines indicate the possible paths in the phase diagrams induced by the dependence of the interaction parameters on the gate distance $d$ shown in Fig.\,\ref{fig:U0UQ}. {(a)} The spinless $N_{\text{s}} = 1$ case. {\bf (b)} The spinful $N_{\text{s}} = 2$ case. {\bf (c-d)} At $N_{\text{s}} = 2$, distributions of the complex order parameter $\langle \hat{M}_{\boldsymbol{K}_{\text{M}}}\rangle = \mbox{Tr}\,\hat{G} \hat{M}_{\boldsymbol{K}_{\text{M}}}$ for $Q / t, U / t = (84.4, 0.24)$ and $(96.6, 0.24)$, respectively. These parameters are marked as stars in panel {\bf (b)}. The distributions are characteristic for the {\bf (c)} cVBS and {\bf (d)} pVBS Kekulé orders. }
    \label{fig:phase_diagram}
\end{figure*}

We consider a model with two orbitals on each moiré honeycomb lattice site representing the two valleys of the microscopic graphene lattice. The dynamics of these orbitals is given by a (nearest-neighbor) hopping with strength $t$, which sets the energy scale. We study both the case of a single spin species $N_{\text{s}} = 1$ as well as the case of two spin species $N_{\text{s}} = 2$, relevant to studying phases with non-trivial pseudo-spin structure. 
We study the model at half-filling. The physical system always has spin, and the case $N_{\text{s}} = 2$ at half-filling corresponds to studying the charge-neutrality point $\nu = 0$ of the physical system. The model with $N_{\text{s}} = 1$ is relevant for the spin-polarized phases of the physical system at $\nu = \pm1$. A spin-polarized phase at $\nu=-2$ only has one spin species occupied and the other spin-species is at half-filling, therefore $N_{\text{s}} = 1$ at half-filling is the relevant case to study. On the other hand, a spin-polarized phase at $\nu=+2$ has one spin species fully occupied (and therefore inert) while the other spin species is at half-filling. Again, the model with $N_{\text{s}}=1$ at half-filling is relevant.
We observe numerous phases in the $(Q, U)$ phase diagram, including the weakly-interacting Dirac semimetal phase (DSM) Kekulé VBS, Néel and charge-density wave (CDW). 

Figure~\ref{fig:phase_diagram} summarizes the phase diagram we obtain within Quantum Monte Carlo (QMC)~\footnote{Note that from Fig.\,\ref{fig:phase_diagram}, one can see that the transitions between these phases take place in the regions of seemingly very large $Q / t$. Importantly, these values are accessible within the real TBlG materials, as we point out in the Section\,\ref{section:model}. Additionally, this is also due to the cluster interaction definition: a violation of a single cluster bracket in Eq.\,\eqref{eq:hamiltonian} would lead to an energy gap of $\Delta_Q = Q / 9$, which should be considered as the characteristic interaction energy scale. Therefore, at the transition point, $\Delta_Q \sim 6 t$, which lies in the hard {\it intermediate-interaction regime}, where the interaction energy scale is comparable to the free tight-binding Hamiltonian band width. This is both far from the free theory and the Mott insulator case, which requires application of an unbiased exact QMC approach.}. The DSM phase is characterized by vanishing susceptibilities and single-particle gap.
In the $N_{\text{s}} = 1$ case, the three valley Néel orders $\tau^v$ are present, where $\tau^v$ are the Pauli matrices with $v \in\{1,2,3\}$ denoting the valley degree of freedom. These order parameters form the adjoint representation of the $SU(2)$ group and their susceptibilities are degenerate. The phase diagram is shown in Fig.\,\ref{fig:phase_diagram}\,(a), where we depict the $\tau^3$ order. In turn, in the Kekulé state, namely columnar VBS (cVBS) emerging in the moiré scale, bond singlets and plaquettes are formed. At the same time, in the $N_{\text{s}} = 2$ setup, we do not observe a Néel state, but rather two Kekulé states, cVBS and plaquette VBS (pVBS). The phase diagram is shown in Fig.\,\ref{fig:phase_diagram}\,(b).

In the valley Néel phase, where three degenerate orders are expected, we consider how the possible perturbations to the Hamiltonian could break the degeneracy. We find that the leading perturbation, next-to-nearest-neighbor hopping, favors the intervalley orders such as $\tau^v$ with $v \in \{1, 2\}$. We also show that this phase has distinct features in the local electron density observable in scanning tunneling microscope (STM) experiments. 

Importantly, the strong cluster interaction $Q$ fixes the number of electrons to $6 N_{\text{s}}$ {\it per each hexagon}, which we call the {\it cluster rule}, and the states satisfying this constraint --- the {\it cluster-rule manifold}. Since the hexagons share corners, the cluster-rule manifold cannot be decomposed into a product of simple local Hilbert spaces, similarly to the iconic pyrochlore spin ice~\cite{Hermele_2004}, where the total magnetization per a (corner-sharing) tetrahedron is constrained to zero. This is in contrast to the case of the strong on-site repulsion $U$, which limits the occupation to one fermion per site and turns the Hubbard model into the Heisenberg model with simple on-site degrees of freedom. Within this cluster-rule manifold, we develop an intuitive perturbation theory and supplement the QMC results with exact diagonalization for $N_{\text{s}} = 1$.
 
The paper is organized as follows. In Section~\ref{section:model}, we introduce the model and the relevant interactions. In Section~\ref{section:results}, we study the model within several numerical techniques, address breaking degeneracy between Néel states and show possible STM images. Finally, in Section~\ref{section:discussion} we discuss the obtained results.

\section{Model}
\label{section:model}
The Wannierization of TBlG conducted in Ref.\,\onlinecite{PhysRevX.8.031087} leads to orbitals centered at the sites of a honeycomb lattice on the moiré scale. The sites of the honeycomb lattice represent the AB and BA stacked regions of the bilayer. Similarly to Ref.\,\onlinecite{PhysRevX.8.031087}, we consider the fermionic model on the honeycomb lattice with two orbitals (valleys) and $N_{\text{s}}$ spins, with $N_{\text{s}} = 1$ or $2$. In the case of TBlG, the orbitals are delocalized, thus an electron on a site has its wave function density concentrated in the three {\it blobs} located in the centers of the three hexagons adjacent to the site, which correspond to the AA/BB regions of the bilayer. For simplicity, we consider a real nearest-neighbor kinetic term with hopping strength $t$, which yields the $SU(2 N_{\text{s}})$--symmetric Hamiltonian

\begin{gather}
    \label{eq:hamiltonian}
    \hat{H} = t \sum\limits_{\langle i,\,j\rangle} \sum\limits_{\sigma, \tau} \hat{c}^{\dagger}_{i \sigma  \tau} \hat{c}^{\phantom{\dag}}_{j \sigma \tau} + \frac{Q}{2} \sum\limits_{\varhexagon} (\hat{n}_{\varhexagon} - 2 N_{\text{s}})^2 \\+ \frac{U}{2} \sum\limits_i (\hat{n}_i - N_{\text{s}})^2,\nonumber
\end{gather}
where $\langle i,\,j \rangle$ denote nearest-neighbor sites, $\sigma$ enumerates the $N_{\text{s}}$ spin species and $\tau=+,-$ enumerates the two valleys. The operator $\hat{n}_i = \sum_{\sigma,\,\tau} \hat{c}^{\dagger}_{i\sigma\tau} \hat{c}^{\phantom{\dag}}_{i\sigma\tau}$ measures the full charge at a site and the hexagon charge operator is given by $\hat{n}_{\varhexagon} = (1/3) \sum_{i \in \varhexagon} \hat{n}_i$. The factor $(1/3)$ stems from the delocalization of Wannierized fermionic orbitals into three peaks.

\begin{figure}[t!]
    \centering
    \includegraphics[width=\columnwidth]{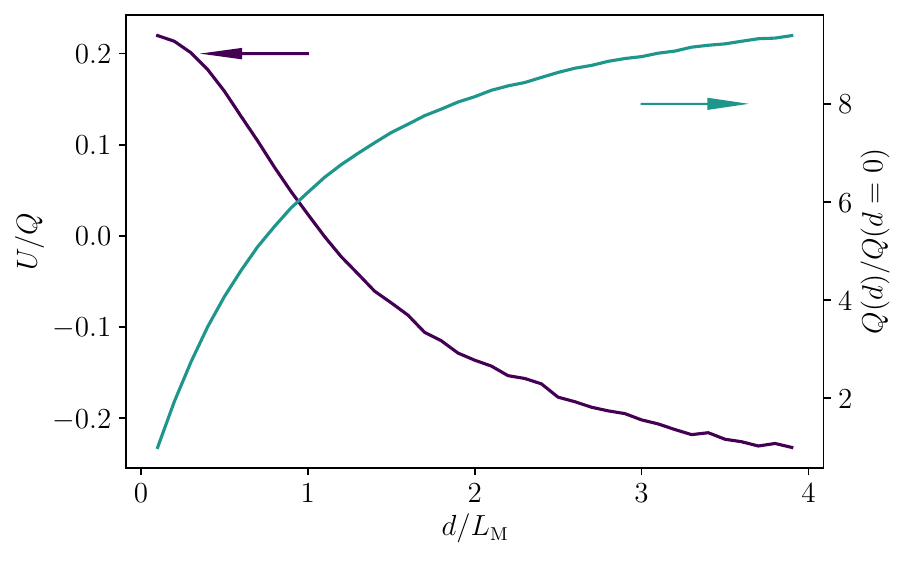}
    \caption{{\bf (Left axis)} Ratio of the on-site potential correction $U$ to the cluster interaction $Q$ as a function of $d / L_{\text{M}}$, where $d$ is the distance to the metallic gates from the TBlG sample. {\bf (Right axis)} Dependence of the screened cluster interaction $Q(d)$ on the gate distance $d$. The details of both calculations are given in Appendix\,\ref{appendix:potentials_shape}.}
    \label{fig:U0UQ}
\end{figure}

The estimates for the characteristic Coulomb energy $Q = e^2 / (0.28 L_{\text{M}}) \sim 220\,\text{meV}$ (here, $L_{\text{M}}$ is the moiré length scale) and the bandwidth $\Delta \sim 6 t \sim 2 \,\text{meV}$ given in Ref.\,\onlinecite{PhysRevX.8.031087} for $\theta \sim 1.07^{\circ}$ suggest that the ratio $Q / t$ can reach hundreds. The cluster interaction depends on the gate distance $d$, as shown in Fig.\,\ref{fig:U0UQ} in agreement with Ref.\,\onlinecite{Stepanov_2020}.

In the absence of the on-site interaction term $U$, the cluster interaction can be seen as an extended Hubbard model with the long-range repulsive contributions decaying as $V_0/V_1/V_2/V_3 = 3/2/1/1$. These ratios are obtained in Ref.\,\onlinecite{PhysRevX.8.031087}, where the interaction strength stems solely from the number of overlapping wave function density centers. In the experimental setup with the interaction screened by two parallel metallic gates placed at the distance $d$ to the TBlG sheet, this cluster-interaction limit corresponds to the case $d \ll L_{\text{M}}$, where $L_{\text{M}}$ is the TBlG moiré lattice spacing~\cite{PhysRevX.8.031087,Bernevig_2021}. When the metallic gates are moved away from the TBlG sheet, the cluster interaction is supplemented with the screened non-local Coloumb interaction between the non-overlapping blobs. This modification of the cluster interaction violates the $3/2/1/1$ extended interactions ratio, which we mimic by introducing the on-site interaction $U$. In Fig.\,\ref{fig:U0UQ}, we show the ratio $U/Q$ as a function of $d / L_{\text{M}}$. Importantly, in an experiment with metallic gates, both positive and negative values of $U / Q$ can be realized. In Fig.\,\ref{fig:phase_diagram}\,(a-b), we draw the possible lines that can be induced by the dependence of $Q(d) / Q(d = 0)$ and $U(d) / Q(d)$ on $d$. The fully-screened interaction strength $Q(d = 0) / t$ depends on the twist angle $\theta$. Notably, these lines pass through all phases, making the variation of $d$ a suitable way to access the observed symmetry-brkoen phases experimentally.


\section{Results}
\label{section:results}

\subsection{Order parameters}
In this Section, we sketch the order parameters emerging in the phase diagram of the Hamiltonian Eq.\,\eqref{eq:hamiltonian}.

\subsubsection{Kekulé valence-bond solids}
\label{subsec:orders}
In the Kekulé VBS phase, reported previously in Ref.\,\cite{qmc_paper}, the order is given by the operator
\begin{gather}
\label{eq:kekule}
    \hat{M}_{\pm \boldsymbol{K}_{\text{M}} } = \frac{1}{\sqrt{N_{\text{b}}}} \sum\limits_{\xi, i, \rho_j} e^{\pm i \boldsymbol{K}_{\text{M}} \boldsymbol{r}_i} \left(\hat{c}^{\dagger}_{i,\xi} \hat{c}_{i + \rho_j, \xi} e^{2 \pi i j / 3} + \text{h.c.}\right),
\end{gather}
where $N_{\text{b}} = 3 L^2$ is the number of bonds in the lattice, $\rho_i$ enumerates the nearest neighbors of site $i$, and $\xi \in \{0, \dots, 2 N_{\text{s}} - 1\}$ labels the flavors. This order transforms as the $E$ representation of the $C_3$ group and has the spatial momenta $\boldsymbol K_{\text{M}}$ and $\boldsymbol K_{\text{M}}'$, the Dirac points in the moiré Brillouin zone. 

The matrix representations $M_{\alpha \beta}$ of the $\pm \boldsymbol K_{\text{M}}$ Kekulé orders Eq.\,\eqref{eq:kekule} are related by complex conjugation and have degenerate susceptibilities. In Ref.\,\onlinecite{qmc_paper}, it was shown that these two orders condense in two real-valued superpositions $(\hat{M}_{+ \boldsymbol K_{\text{M}}} + \hat{M}_{- \boldsymbol K_{\text{M}}}) / \sqrt{2}$ and $(\hat{M}_{+ \boldsymbol K_{\text{M}}} - \hat{M}_{- \boldsymbol K_{\text{M}}}) / (i \sqrt{2})$, the pVBS and cVBS, respectively.  

The determination of which phase is realized can be done by plotting the histogram of the $\langle \hat{M}_{+ \boldsymbol K_{\text{M}}} \rangle$ measurements within QMC in the complex plane. The characteristic distributions of the order parameter $\langle \hat{M}_{+ \boldsymbol K_{\text{M}}} \rangle$ are shown in Fig.\,\ref{fig:phase_diagram}\,(c-d).

\subsubsection{Néel antiferromagnets}

In the Néel AFM phase, the symmetry breaking may take place due to the condensation of the orders
\begin{gather}
\label{eq:Neel}
\hat{M}^{v, s} = \frac{1}{2 \sqrt{2 L^2}} \sum\limits_{i} (-1)^{l(i)} \left(\boldsymbol{\hat c}^{\dagger}_{i} t^{v, s} \boldsymbol{\hat c}^{\phantom{\dag}}_{i} + \text{h.c.} \right),
\end{gather}
where $\hat{\boldsymbol{c}}^{\dagger}_{i}$ is the collection of $2 N_{\text{s}}$ creation operators at the site $i$, $l(i)=0,1$ is the sublattice index of the site $i$, and $t^{v, s}$ is one of the $(2 N_{\text{s}})^2 - 1$ Lie algebra generators of the $SU(2 N_{\text{s}})$ group. Here, in the $N_{\text{s}} = 2$ case, we choose the standard representation 
\begin{gather}
    t^{v, s} = \sigma^s \otimes \tau^v,
\end{gather}
where $0 \leqslant v, s < 4$ enumerate the Pauli matrices with $v + s > 0$. For the $N_{\text{s}} = 1$ case, $t^{v} = \tau^v$ with $v > 0$.

\subsubsection{Charge-density wave}
Lastly, in the CDW the order is described by condensation of
\begin{gather}
\label{eq:CDW}
    \hat{M}_{\text{CDW}} = \frac{1}{L} \sum\limits_i (-1)^{l(i)}\hat{n}_{i}.
\end{gather}

\subsection{Quantum Monte Carlo study}
\label{subsec:QMC}
The Hamiltonian Eq.\,\eqref{eq:hamiltonian} can be studied within the Quantum Monte Carlo (QMC) approach~\cite{qmc_paper}, which is sign-problem free at $U,\,Q > 0$. At $U < 0$, the approach is only sign-problem free at $Q = 0$, which hinders the study of the phase diagram for negative $U$.

We consider the equilateral $L \times L$ clusters with $L~=~3,\,6\,9\,12$ and $15$ and vary the Trotter step $t\,\delta\tau~=~\text{min}\,\left(0.1,\,t/Q, t/U \right).$ This choice of $\delta \tau$ allows us to keep the Trotter errors under control even at strong interaction~\footnote{We use the symmetric Trotter decomposition, where Trotter errors scale as $\propto \delta \tau^2 \|[\hat{T}, \hat{Q} + \hat{U}]\|^2$ with $[\hat{T}, \hat{Q} + \hat{U}]$ the commutator between the kinetic and interaction terms. We set $\delta \tau$ such that the relative systematic energy errors caused by Trotterization are below $10^{-3}$.}. The 4-valued Hubbard-Stratonovich field is introduced using the 4-th order $\mathcal{O}(\delta \tau^4)$ decomposition with the errors negligible as compared to the Trotterization errors~\cite{Assaad_2022}. 

\begin{figure}[t!]
    \centering
    \includegraphics[width=\columnwidth]{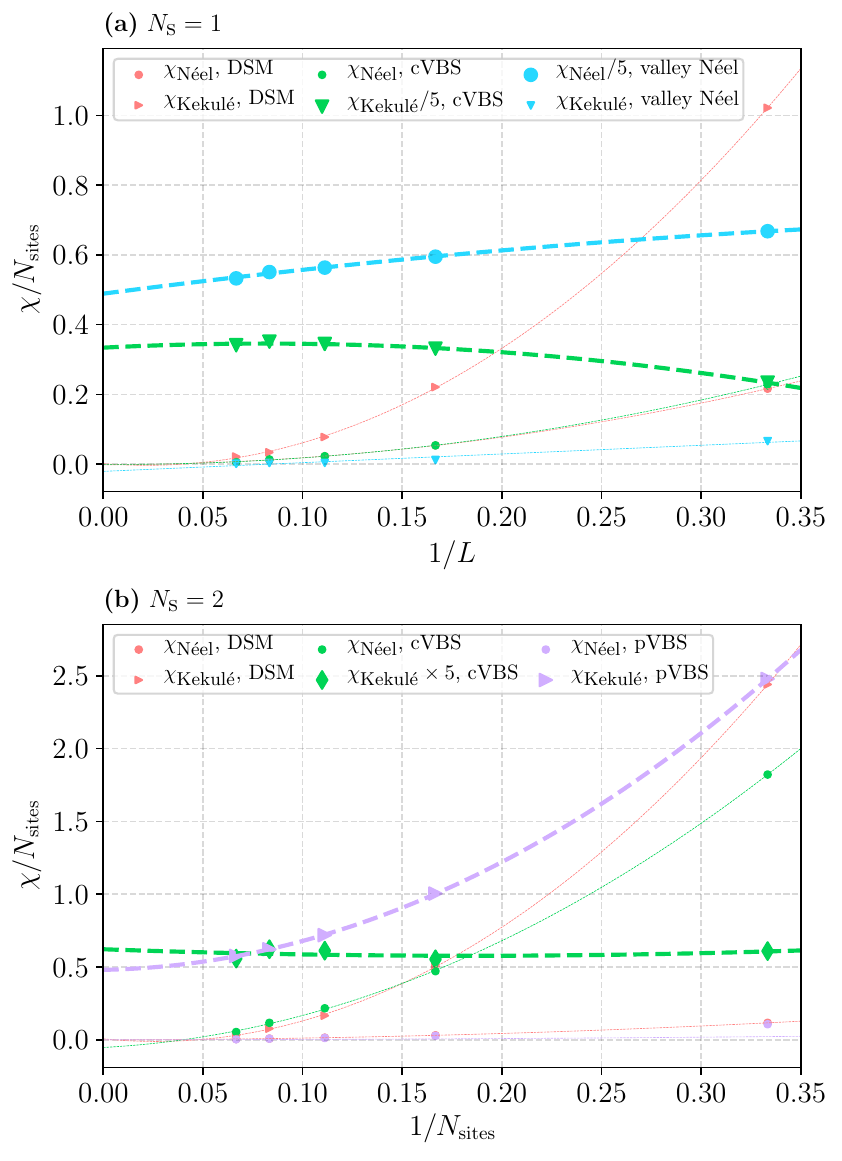}
    \caption{{\bf (a)} $N_{\text{s}} = 1$, quadratic extrapolations of susceptibilities in the DSM, cVBS and valley Néel phases corresponding to the parameters $(U / t, Q / t) = (0, 20),\, (0, 100)$ and $(2, 80)$, respectively. The Kekulé susceptibility in the cVBS phase and the Néel susceptibility in the valley Néel phase are divided by $5$ at $N_{\text{s}} = 1$, while the Kekulé susceptibility in the cVBS phase as multiplied by $5$ at $N_{\text{s}} = 2$ for better general visibility. The non-zero extrapolations are denoted with bold lines. {\bf (b)} Extrapolations in the $N_{\text{s}} = 2$ case in the DSM, cVBS and pVBS phases corresponding to the parameters $(U / t, Q / t) = (0, 40),\, (0, 60)$ and $(6, 60)$, respectively.}
    \label{fig:extrapolations_QMC}
\end{figure}

A generic order parameter has the form
\begin{gather}
    \hat{M} = \sum\limits_{\alpha,\beta} M_{\alpha \beta} \hat{c}^{\dagger}_{\alpha} \hat{c}^{\phantom{\dag}}_{\beta},
\end{gather}
where $\alpha = (i_{\alpha}, \tau^{\alpha}, s^{\alpha})$ is the composite site-flavor index. This operator is normalized as $M^*_{\alpha \beta} M_{\beta \alpha} = 1.$

The tendency to establish a non-zero value of the order $\langle \hat{M} \rangle \neq 0$ can be quantified by the zero-frequency susceptibility 
\begin{equation}
    \begin{split}
    \chi_{\hat{M}} &= \int\limits_{\tau = 0}^{\beta}\mbox{d}\tau\, \mbox{Tr}\, \left[e^{-\beta \hat{H}} \hat{M}^{\dagger}(\tau) \hat{M}(0)\right]\\ 
    &= \sum_{\alpha \beta \gamma \delta} M^*_{\alpha \beta} \Gamma_{\alpha \beta \gamma \delta} M^{\phantom{*}}_{\gamma \delta},
\end{split}
\end{equation}
where we singled-out the four-point particle-hole vertex operator
\begin{gather}
\label{eq:vertex}
    \Gamma_{\alpha \beta \gamma \delta} = \int\limits_{\tau = 0}^{\beta} \mbox{d}\tau\, \mbox{Tr}\, \left[e^{-\beta \hat{H}} \hat{c}^{\dagger}_{\beta}(\tau) \hat{c}^{\phantom{\dag}}_{\alpha}(\tau) \hat{c}^{\dagger}_{\gamma}(0) \hat{c}^{\phantom{\dag}}_{\delta}(0)\right].
\end{gather}

Treating $(\gamma, \delta)$ and $(\alpha, \beta)$ as the composite in- and out-indices, we view $\Gamma_{(\alpha \beta), (\gamma \delta)}$ as a matrix, whose eigenvectors are the operators transforming as irreducible representations of the translation and point-group symmetries and the eigenvalues are the susceptibilities. The irreducible representation of the eigenvector $\hat{M}$ with non-zero extrapolated susceptibility $\lim\limits_{L \to \infty} \chi(L) / L^2$ signals the preferred symmetry-breaking pattern.

For all parameters studied in this work, we find that the leading eigenvector is always one of the orders defined in Section\,\ref{subsec:orders}. Moreover, we reproduce the expected $((2 N_{\text{s}})^2 - 1)$--fold and $2$--fold degeneracies for the $\hat{M}^{v, s}$ Néel and the $\hat{M}_{\pm \boldsymbol K_{\text{M}}}$ Kekulé orders, respectively~\footnote{The full $\Gamma_{\alpha \beta \gamma \delta}$ vertex requires $\mathcal{O}(L^8)$ measurements. We restrict the possible order matrices $M_{\alpha \beta}$ to contain elements only up to the nearest-neighbor, reducing the measurements count to $\mathcal{O}(L^4)$. We check that the results are insensitive to this reduction.}.

We perform QMC simulations at $\beta t = 2 L$ and check for selected parameters that the conclusions are temperature-insensitive at $\beta t = 3 L$. The resulting QMC phase diagrams are shown in Fig.\,\ref{fig:phase_diagram}\,(a-b). 
In Fig.\,\ref{fig:extrapolations_QMC}\,(a) we show the susceptibilities' extrapolations in the $N_{\text{s}} = 1$ case. Here, the extrapolated Néel susceptibility is finite for $(U/t, Q/t) = (2, 80)$, indicating the valley AFM phase, while the Kekulé susceptibility is finite in the cVBS phase for $(U/ t, Q / t) = (0, 100)$. In the DSM phase at $(U / t, Q / t) = (0, 20)$, all susceptibilities extrapolate to zero. In Fig.\,\ref{fig:extrapolations_QMC}\,(b), where $N_{\text{s}} = 2$ is shown, the Kekule susceptibilities are finite in the pVBS phase at $(U/ t, Q / t) = (0, 60)$ and in the cVBS phase at $(U/ t, Q / t) = (6, 60)$, while they extrapolate to zero in the DSM phase at $(U / t, Q / t) = (0, 40)$.

\subsection{Perturbation theory}
\label{subsec:perturbation_theory}
The large $Q/t$ interaction strengths of the phase transitions in Fig.\,\ref{fig:phase_diagram}\,(a-b) suggest construction of an effective theory in the limit of the strong cluster interaction ($Q \gg t, U$), where the low-energy manifold is described by the states having exactly $6 N_{\text{s}}$ fermions in each hexagon.  Notably, unlike the construction of the $t$--$J$ model in the regime of strong $U$, here the resulting cluster-rule manifold can not be written as a product of Hilbert spaces of some emergent local degrees of freedom, which is similar to the pyrochlore lattice~\cite{Hermele_2004}.

The cluster-rule manifold is separated by the gap $\Delta_Q = Q / 9$ from the rest of the states in the full Hilbert space. Within this manifold, the on-site term $U$ is diagonal, while the kinetic term $t$ can be treated perturbatively

\begin{figure}[t!]
    \centering
    \includegraphics[width=\columnwidth]{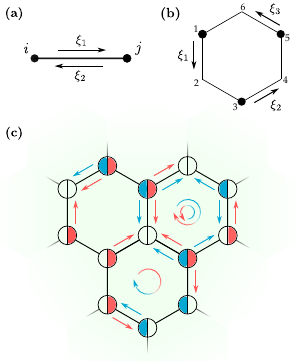}
    \caption{Two moves in the cluster-rule perturbation theory. {\bf (a)} Bond flip exchanging fermions of flavors $\xi_1$ and $\xi_2$ on sites $i$ and $j$. {\bf (b)} The hexagon flip process that moves three fermions with the flavors $\xi_1, \xi_2, \xi_3$ over three non-adjacent bonds in a hexagon $\varhexagon$ in the same direction, such that the cluster rule is respected. {\bf (c)} An example of a cluster-rule configuration for the $N_{\text{s}} = 1$ case. Left (right) colored semi-circles indicate presence of a spin up (down). Note that in each hexagon, there are exactly $6$ fermions. The arrows along the bonds indicate the possible single-particle hops between adjacent sites that are allowed by the Pauli principle. The round arrows in the centers of the hexagon indicate possible {\bf (b)}--moves. Namely, there are two {\bf (b)}--moves possible in the upper-right hexagon, one {\bf (b)}--move possible in the lower hexagon, and no moves possible in the upper-left hexagon.}
    \label{fig:pert_processes}
\end{figure}

\begin{gather}
    \label{eq:perturbative}
    \hat{H}_{\text{pert.}} = (\mathbbm{1} - \hat{\mathcal{P}}) \left(-\hat{K} \frac{\hat{\mathcal{P}}}{\hat{Q}} \hat{K} + \hat{K} \frac{\hat{\mathcal{P}}}{\hat{Q}} \hat{K}\frac{\hat{\mathcal{P}}}{\hat{Q}} \hat{K} + \ldots\right) (\mathbbm{1} - \hat{\mathcal{P}}),
\end{gather}
where $\hat{K}$, $\hat{Q}$ are the kinetic and cluster terms in Eq.\,\eqref{eq:hamiltonian}, and $\hat{\mathcal{P}}$ projects onto the orthogonal complement of the cluster-rule manifold. In this manifold, the effective Hamiltonian reads:
\begin{gather}
    \label{eq:effective_terms}
    \hat{H}_{\text{pert.}} = -\frac{t^2}{\Delta_Q} \hat{h}_{\text{bond}} + \frac{t^3}{\Delta_Q^2} \hat{h}_{\text{hexagon}} + \frac{U}{2} \sum\limits_i (\hat{n}_i - N_{\text{s}})^2.
\end{gather}
Here, the first term,
\begin{gather}
    \hat{h}_{\text{bond}} = \sum\limits_{\xi_1, \xi_2}\sum\limits_{\langle i, j\rangle} \hat{c}^{\dagger}_{\xi_1, i} \hat{c}_{\xi_1, j} \hat{c}^{\dagger}_{\xi_2, j} \hat{c}_{\xi_2, i}
\end{gather}
exchanges the fermions with the flavors $\xi_1, \xi_2$ on the nearest-neighbor sites $i, j$, which is shown in Fig.\,\ref{fig:pert_processes}\,(a). The second term
\begin{gather}
    \hat{h}_{\text{hexagon}} = \sum\limits_{\xi_1, \xi_2, \xi_3}\sum_{\varhexagon} \hat{c}^{\dagger}_{\xi_1, 1} \hat{c}_{\xi_1, 2} \hat{c}^{\dagger}_{\xi_2,3} \hat{c}_{\xi_2, 4}
    \hat{c}^{\dagger}_{\xi_3, 5} \hat{c}_{\xi_3, 6},
\end{gather}
moves three fermions with the flavors $\xi_1, \xi_2, \xi_3$ over three non-adjacent bonds in a hexagon $\varhexagon$ in the same direction, as shown in Fig.\,\ref{fig:pert_processes}\,(b). Note that each of these terms preserves the cluster rule in all hexagons. These terms annihilate a state if a move is not possible. To illustrate these moves, in Fig.\,\ref{fig:pert_processes}\,(c) we show a basis element satisfying the cluster rule. The figure illustrates the dynamics induced by the terms $\hat{h}_{\text{bond}}$ and $\hat{h}_{\text{hexagon}}$ within the cluster-rule manifold.

\subsection{Exact diagonalization of $\hat{H}_{\text{pert.}}$}
\label{subsec:exact_diagonalization}
We treat the Hamiltonian Eq.\,\eqref{eq:perturbative} within exact diagonalization (ED)~\footnote{In principle, one could consider the $3 \times 3$ system with $N_{\text{s}} = 1$ directly in the form of the initial Hamiltonian Eq.\,\eqref{eq:hamiltonian}. However, this strategy would have a significant drawback: due to the small system circumference, the perturbative expansion in the form Eq.\,\eqref{eq:perturbative} would produce, in addition to the terms present in Eq.\,\eqref{eq:perturbative}, non-physical third-order processes stretching over the system boundary, that are absent in larger systems.} in the $N_{\text{s}} = 1$ case on the $3 \times 3$ hexagonal lattice. The cluster-rule manifold contains $5`018`802$ states, which is a considerable reduction from the full Hilbert space with $9 + 9$ fermions of the size $\sim 3 \times 10^{9}$. Unfortunately, the cluster-rule manifold on the $3 \times 3$ lattice with $N_{\text{s}} = 2$ has the size of $\mathcal{O}(10^{13})$ and is out of reach. 

\begin{figure}[t!]
    \centering
    \includegraphics[width=\columnwidth]{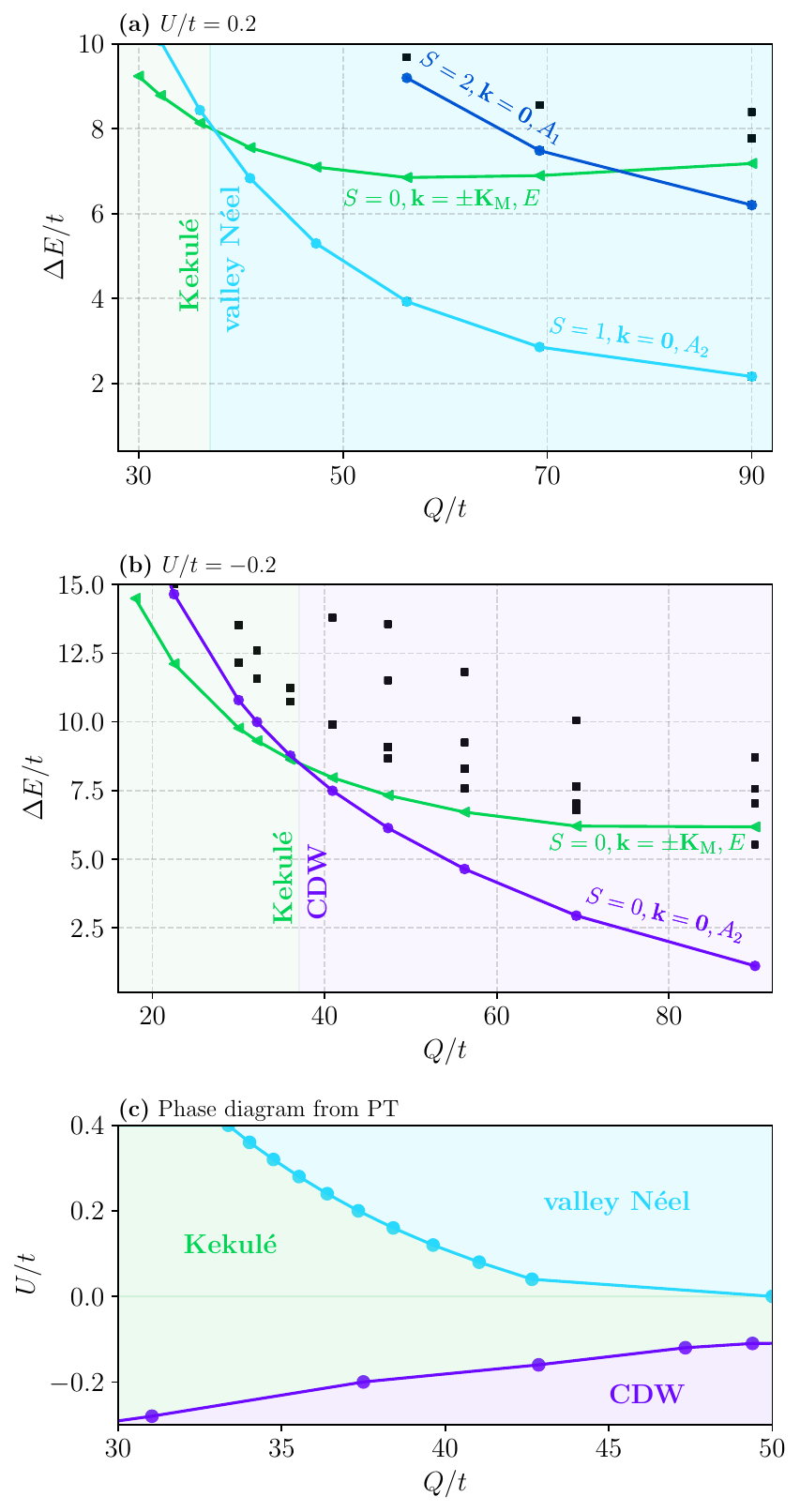}
    \caption{{\bf (a)} Spectrum obtained within exact diagonalization of the Hamiltonian Eq.\,\eqref{eq:perturbative} at $U / t = 0.2$. Here, we plot the energy gap between an excitation (and mark its quantum numbers) and a featureless ground state (with quantum numbers $S = 0$, $\boldsymbol k = 0,$ $A_1$). The lowest excitations in selected symmetry sectors are connected with lines, and their quantum numbers are given. Other excitations are shown as simple black squares. {\bf (b)} The case $U / t = -0.2$. {\bf (c)} The resulting phase diagram with phase transitions pinned from the crossings of the excited states' energies.}
    \label{fig:ed_spectrum}
\end{figure}

Figure~\ref{fig:ed_spectrum}\,(a) shows the low-energy level structure at $U / t = 0.2$ as a function of $Q / t$. We determine the phase diagram on the $Q / t$ axis by considering the quantum numbers of the low-energy excited states, namely their spin $S$, momentum $\boldsymbol k$ and their transformation property under the point group $D_3$. In the regime of small $Q / t$, the lowest excitations above the (featureless) ground state, are singlets $S = 0$ with $\boldsymbol k = \boldsymbol K_{\text{M}} / \boldsymbol K_{\text{M}}'$, transforming as the $E$ representation under the $D_3$ group, which matches the symmetry-breaking pattern of the Kekulé state. In the large--$Q / t$ regime, we observe the magnetic tower of states signalling breaking of the $SU(2)$ symmetry and establishing a Néel order~\cite{wietek2017studying}.
Following Refs.\,\onlinecite{Suwa_2016,imadanomura}, we identify the transition between the magnetic Néel phase at large $Q / t$ and Kekulé phase at small $Q / t$ with the crossing between the lowest-lying triplet and the symmetry-breaking singlet Kekulé state.

Negative $U / t = -0.2$ favours doubly-occupied or free sites. There, the phase transition is pinned by the crossing between the Kekulé and the CDW $(S = 0, \boldsymbol k = \boldsymbol 0, \,A_2)$ excited states, signalling the $\boldsymbol k = \boldsymbol 0$ CDW, as we show in Fig.\,\ref{fig:ed_spectrum}\,(b). With these transition criteria, in Fig.\,\ref{fig:ed_spectrum}\,(c) we show the resulting phase transition lines between the Kekulé and the CDW phases. At $U > 0$ the results qualitatively agree with the QMC phase diagram shown in Fig.\,\ref{fig:phase_diagram}\,(a). The actual interaction strengths differ due to the significant finite-size effects present in $L = 3$ perturbation theory. Importantly, the region of $U < 0$ is not accessible for the exact QMC study due to a severe sign problem.

\subsection{Fixed-point states analysis}
\label{subsec:wave_functions}
The absence of a Néel phase for $N_{\text{s}} = 2$, which is in contrast to the $N_{\text{s}} = 1$ case, can be explained by constructing the fixed-point states of the orders observed within QMC and ED. The corresponding orders are depicted in Fig.\,\ref{fig:phase_diagram}\,(a-b). A fixed-point state is obtained as a ground state of the Hamiltonian $\hat{H} = \hat{M}$, where $\hat{M}$ is one of the order parameters, i.\,e., the condensed order dominates the initial Hamiltonian. Finally, we project this state onto the cluster-rule manifold by removing the wave function components violating the cluster rule.

The possible Néel $SU(4)$ state (for simplicity, we consider the $\sigma^0 \otimes \tau^3$ order) reads
\begin{gather}
    \label{eq:fpwf_Neel}
    |\psi_{\text{Néel}}\rangle = \left(\prod\limits_{i=1}^{L^2} \prod\limits_{s = 1}^{2} \hat{c}^{\dagger}_{i, A, s, +} \hat{c}^{\dagger}_{i, B, s, -}\right) |0\rangle,
\end{gather}
where the first (second) operator creates fermions in the $\tau = +1 (-1)$ valley on the $A(B)$ sublattice. In the $SU(2)$ spinless case, the spin degree of freedom in Eq.\,\eqref{eq:fpwf_Neel} is omitted.  

For the cVBS phase, the state reads
\begin{gather}
    \label{eq:pfwf_cVBS}
    |\psi_{\text{cVBS}}\rangle = \frac{1}{(2 N_{\text{s}})^{N_{\text{b}} / 3}}\prod\limits_{b = 1}^{N_{\text{b}} / 3} \sum\limits_{c=0}^{2 N_{\text{s}} \choose N_{\text{s}}} |v_c^{\alpha_{b}}\rangle \otimes |\bar{v}_c^{\beta_{b}}\rangle,
\end{gather}
where $b$ enumerates the $N_{\text{b}} / 3$ dimerized bonds between sites $(\alpha_{b}, \beta_{b})$, while $c$ enumerates $2 N_{\text{s}} \choose N_{\text{s}}$ possible ways to put $N_{\text{s}}$ fermionic species at the site $\alpha_{b}$ and the remaining $N_{\text{s}}$ species at the site $\beta_b$, forming the state $|v_c^{\alpha_{b}}\rangle \otimes |\bar{v}_c^{\beta_{b}}\rangle$. This state has exactly $N_{\text{s}}$ electrons per site and thus satisfies the cluster rule.

We compute the energies of these fixed-point states according to Hamiltonian Eq.\,\eqref{eq:perturbative}, yielding
\begin{gather}
    E_{\text{Néel}} \Delta_Q / t^2 = -2 N_{\text{s}} N_{\text{b}}, \\
    E_{\text{cVBS}}^{N_{\text{s}} = 1} \Delta_Q / t^2 = -2 N_{\text{b}},\; E_{\text{cVBS}}^{N_{\text{s}} = 2} = -\frac{16}{3} N_{\text{b}}.
\end{gather}

We see that at $N_{\text{s}} = 1$, the Néel and cVBS energies are equal, $E_{\text{Néel}} = E_{\text{cVBS}}^{N_{\text{s}} = 1}$, which opens the prospect to realize the $SU(2)$ valley Néel state in Fig.\,\ref{fig:phase_diagram}\,(a) due to higher-order quantum fluctuations. On the contrary, at $N_{\text{s}} = 2$, $E_{\text{Néel}} > E_{\text{cVBS}}^{N_{\text{s}} = 2}$, and thus the $SU(4)$ Néel state in Fig.\,\ref{fig:phase_diagram}\,(b) is never a ground state~\footnote{Within the QMC simulations, at finite volumes $L = 3$ and $6$, there exist the parameters region at $N_{\text{s}} = 2$ where the $SU(4)$ Néel is the ground state of the vertex Eq.\,\eqref{eq:vertex}. This, however, does not hold in the thermodynamic limit, where the cVBS or pVBS extrapolations are always dominant.}.

\subsection{Breaking the degeneracy between the Néel states}
\label{subsec:breaking_degeneracy}
We observed that in the spinless case $N_{\text{s}} = 1$, the valley Néel phase appears in the phase diagram. The three Néel states are degenerate within the given $SU(2)$--symmetric model Eq.\,\eqref{eq:hamiltonian}. However, additional perturbations in the Hamiltonian can break this symmetry, which may lead to a splitting of the degeneracy between these Néel orders. 

In Eq.\,\eqref{eq:hamiltonian}, we neglected additional long-range hoppings and longer-range interactions emerging due to the delocalized nature of the Wannier orbitals~\cite{PhysRevX.8.031087}. The next-to-leading hopping obtained in Ref.\,\cite{PhysRevX.8.031087} is the fifth-nearest-neighbor hopping with the respective Hamiltonian
\begin{gather}
    \hat{H}_{5} = t_{5} \sum\limits_{\langle\langle i,\,j\rangle\rangle} \hat{c}^{\dagger}_{i +} \hat{c}_{j +} + t_{5}^*\sum\limits_{\langle\langle i,\,j\rangle\rangle}\hat{c}^{\dagger}_{i -} \hat{c}_{j -},
\end{gather}
where $\langle\langle i,\,j\rangle\rangle$ denotes the fifth-nearest-neighbor pairs. Crucially, while the nearest-neighbour hopping $t$ can always be chosen to be real via an appropriate choice of gauge, $t_{5}$ is generally complex~\footnote{Note that our $t_5$ corresponds to $t_2$ in the notation of \cite{PhysRevX.8.031087}.}. With a complex hopping, the symmetry of the model is broken down from valley $SU(2)$ symmetry to $U(1)$ valley charge conservation. 

Since this term hops a single particle, there is no contribution in the first-order perturbation theory. The second-order perturbation theory (including only the imaginary component of the $t_5$ hopping since this is the piece which breaks the symmetry) yields
\begin{figure*}[t!]
    \centering
    \includegraphics[width=\textwidth]{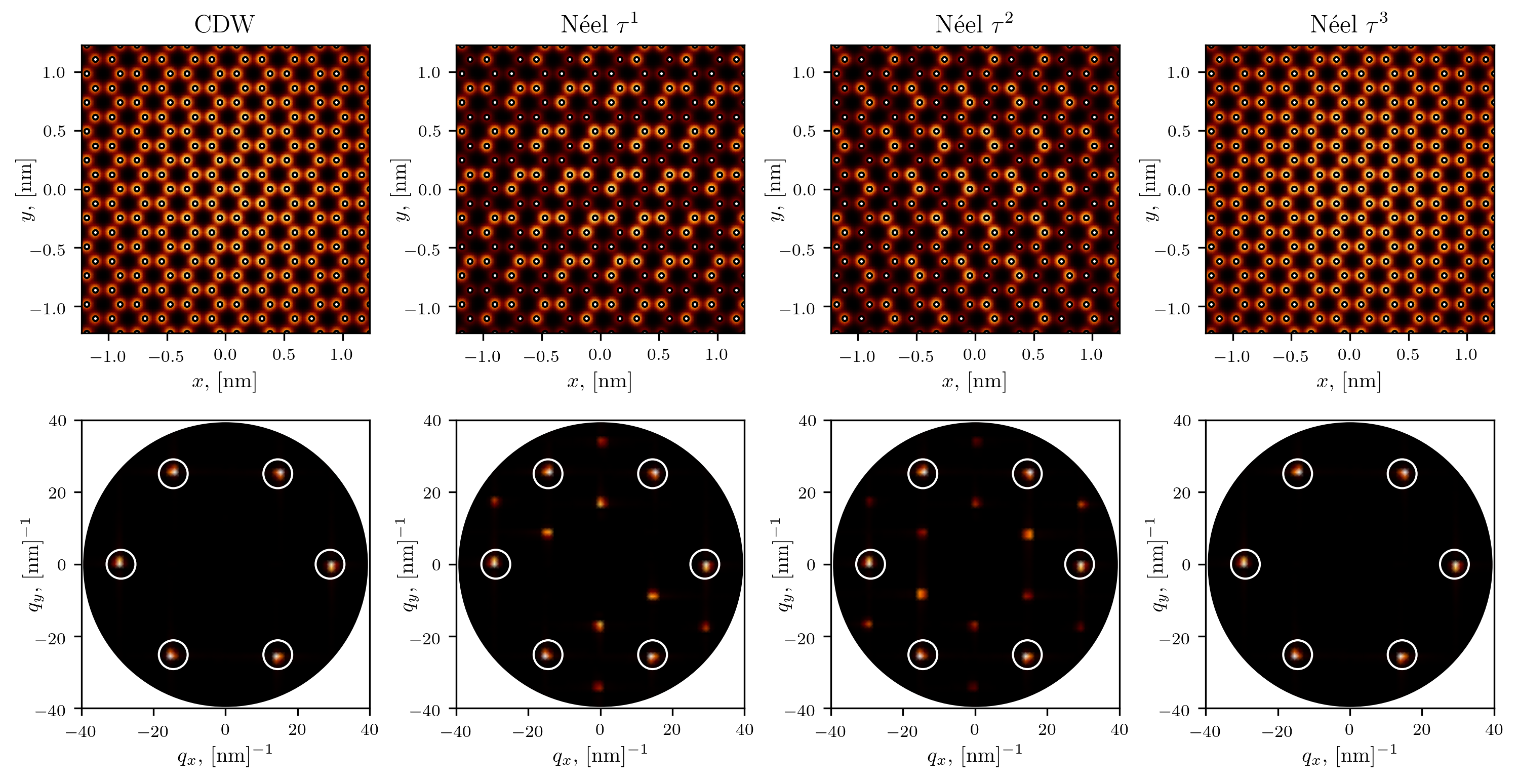}
    \caption{{\bf Upper row:} Local electron density $\rho(\boldsymbol r)$ of the CDW and the $\tau^1$, $\tau^2$ and $\tau^3$ Néel orders, which could be measured using STM. The images show the local electron density in one of the graphene layers on the microscopic graphene scale. White dots show the positions of the sites of one of the graphene layers. {\bf Bottom row:} Fourier transforms $\rho(\boldsymbol q)$ of the respective STM images. The central peak at $\boldsymbol q = \boldsymbol 0$ is removed for visibility. The six graphene Bragg peaks are circled, the additional peaks in the Fourier transform for the $\tau^1$ and $\tau^2$ orders are a result of the $\sqrt{3}\times\sqrt{3}$ translational symmetry breaking in the presence of intervalley coherence and correspond to momenta $\boldsymbol K-\boldsymbol K'$.
    }
    \label{fig:STM}
\end{figure*}
\begin{gather}
    \delta E^{(2)}_{v} = \begin{cases}
                        -\displaystyle \frac{2 |\text{Im}\, t_{5}|^2}{3 \Delta_Q} N_{\text{b}},\;\text{if $v \in (1, 2)$} \\
                        0,\;\text{otherwise}
                    \end{cases}.
\end{gather}
Therefore, adding the 5-th nearest-neighbor hopping favours inter-valley Néel orders, i.\,e.~the system is an easy-plane valley-antiferromagnet.

\subsection{Graphene scale electron density distribution}
\label{subsec:STM}
STM topography experiments have recently revealed that the correlated insulating phases in TBlG~\cite{nuckolls2023quantum} and twisted trilayer graphene~\cite{kim2023imaging} exhibit a Kekulé pattern on the microscopic graphene scale. This is a signature of certain types of intervalley coherent (IVC) states~\cite{STM_1,STM_2} and is consistent with predictions from Hartree-Fock on the continuum model~\cite{IKS_PRX,IKS_PRL,kwan2023electronphonon}. 

Since STM has been shown to be a powerful tool for identifying the nature of symmetry breaking in TBlG, we plot in Fig.\,\ref{fig:STM} for $N_{\text{s}} = 1$ the potential STM images that would appear on the microscopic graphene scale of the bilayers, when the respective orders we found stabilize on the moiré scale. In particular, we show the local electron density in one of the layers for the CDW and for three different Néel orders, all of which break moir\'e translational symmetry. The $\tau^1$ and $\tau^2$ N\'eel orders allow for superposition of electrons from the two microscopic graphene valleys and this leads to a $\sqrt{3}\times\sqrt{3}$ increase of the size of the microscopic graphene unit cell.  In the Fourier transform of the STM images, the graphene-scale translational symmetry breaking shows up as additional peaks besides the graphene Bragg peaks. Using the nomenclature of Ref.\,\onlinecite{nuckolls2023quantum} there can be bond as well as site Kekulé order. Intrasublattice IVC leads to a site Kekulé pattern whereas intersublattice IVC leads to a bond Kekulé pattern. 
In addition, the $\tau^1$ and $\tau^2$ Néel orders both break $C_3$ symmetry. The degeneracy between the different Néel orders is lifted by the $t_5$ term which favours the intervalley Néel order (with a Kekulé pattern).

\section{Discussion and outlook}
\label{section:discussion}
In this work, we considered an extended Hubbard model for twisted bilayer graphene with two valleys and $N_{\text{s}}$ spins on the honeycomb lattice. The topology of the bands of twisted bilayer graphene leads to delocalized Wannier orbitals and therefore the onsite Hubbard interaction $U$ is subdominant. Instead, the dominant interaction term is a special cluster interaction $Q$ within each hexagonal plaquette of the honeycomb lattice. Experimentally, the range of the Coulomb interaction and therefore the competition between $U$ and $Q$ can be tuned by modifying the distance between the sample and the gates. While previous works have focused on this model for $U=0$~\cite{PhysRevX.8.031087,qmc_paper}, we extend the study to non-zero $U$ and show that the presence of this term can stabilize a new phase. We studied the phase diagram of this model in $U$--$Q$ space using QMC. While the QMC is limited to $U>0$ due to the sign problem, we make progress on the $U<0$ side of the phase diagram by performing ED within the manifold of states satisfying a cluster rule for large $Q$. The different competing phases we find include a Dirac semimetal, as well as three symmetry-broken phases (CDW, Néel and VBS). In particular, the Néel phase is only stabilized for nonzero $U$.

We explored both $N_{\text{s}}=1$ and $N_{\text{s}}=2$ and find that the phase diagrams are different: In particular, the Néel phase only appears for $N_{\text{s}}=1$. For $N_{\text{s}}=1$, we find good qualitative agreement between the QMC and the perturbation-theory approaches in the $U>0$ case, where there are solutions from both methods. However, the perturbation theory is too computationally expensive in the $N_{\text{s}}=2$ case and therefore the $U<0$ part of the phase diagram remains unexplored in this case. This would be an interesting regime to study in future work, for example using density-matrix renormalization group methods.

Besides, the cluster rule of having exactly $6 N_{\text{s}}$ fermions per hexagon, while the hexagons share corners, appears very similar to the famous pyrochlore spin-ice rule~\cite{Hermele_2004}, where the tetrahedra are also corner-sharing. There, this 'frustrated' local charge conservation allows to reformulate the model in terms of a frustrated gauge theory hosting a $U(1)$ algebraic spin liquid. A similar $U(1)$ gauge theory can possibly be built in this model with the cluster interaction. However, this $U(1)$ gauge theory would be confined in two dimensions~\cite{zhong2012twodimensional}, in accord with us not finding any spin liquid phases. The construction of this gauge theory is beyond the scope of this work. 

Finally, one of the key questions is what symmetry-breaking phases are realized in the experimental system. Many of the experiments performed on TBlG perform transport measurements and as such are unable to distinguish between the different patterns of symmetry breaking: these states all appear as otherwise featureless insulating states in transport. However, recently advances have been made in using STM in order to image TBlG~\cite{nuckolls2023quantum,kim2023imaging}. This allows one to image the form of translational symmetry breaking that is characteristic of the different phases. We computed the STM pattern for the CDW and Néel phases found in the QMC and showed that they can indeed be distinguished with such a measurement. 

In our work, we studied one of the two main models employed to describe TBlG, namely a Hubbard-like model with fidget spinner Wannier orbitals descending from the symmetries of TBlG~\cite{Kang,PhysRevX.8.031087,Po}. This Hubbard-like model allows us to utilize the local cluster-rule constraints and access the physics of TBlG from the strong-coupling regime. This real-space description stands in contrast to the often-used momentum-space description in terms of the flat bands of the continuum model~\cite{Bistritzer2011} whose strong-coupling ground states are generalized quantum Hall ferromagnets~\cite{Bultinck}. The mechanism behind these ground states is the Stoner ferromagnetism due to the exchange interactions. On the other hand, Hubbard-like models often give rise to antiferromagnetic states such as the Néel state in our model, in a way that can be understood using Anderson superexchange in the simplest case. Which of these two descriptions is most appropriate depends crucially on the range of the interactions as well as the comparison between the exchange and superexchange energy scales. One of the great strengths of TBlG and van-der-Waals materials in general is their tunability, via gating, substrate potentials and twist angle. In our work we showed how this tunability allows access to different phases in the phase diagram of the Hubbard model studied. In future, it may be possible to tune the system between the Hubbard model regime and the continuum model regime.


\section{Acknowledgements}
N.\,A.~is funded by the Swiss National Science Foundation, grant number: PP00P2{\_}176877. We acknowledge funding from the European Research Council (ERC) under the European Union’s Horizon 2020 research and innovation program via ERC-StG-Neupert-757867-PARATOP (T.\,N., G.\,W.). The QMC simulations were performed using the high-performance \texttt{ALF} package~\cite{Assaad_2022}. We thank (\textdagger)\,Alexey A.\,Soluyanov and (\textdagger)\,Sandro Sorella for contributing during the early stages of the project.

\appendix
\section{Computation of potential signatures in STM}
\label{appendix:STM}

Here, we provide information on computing the STM images of the different orders obtained in this work. The Wannierized wave functions centered around $BA$ and $AB$ stacking regions ($A$ and $B$ sublattices in the moiré scale, respectively) can be projected onto the two layers $L = 1, 2$ and their sublattices $l = A, B$. The resulting projections $|\psi_{A/B, L, l}\rangle$ are given in the panels of Fig.\,3 of Ref.\,\cite{PhysRevX.8.031087} (note that in Ref.\,\cite{PhysRevX.8.031087}, $|\psi_{A/B}\rangle$ are named 
 $|\psi_{1/2}\rangle$). 

One can see that generally a wave function contribution to $(L, l)$ can be seen as three blobs concentrated around the $AA$ stacking regions with some phases. To describe a blob concentrated around an $AA$ region with the center $\boldsymbol r_0$, we consider a Gaussian contribution $\phi^{\boldsymbol r_0}(\boldsymbol r) \propto \exp(-|\boldsymbol r - \boldsymbol r_0|^2 / D^2)$ with $D \sim L_{\text{M}} / 3$.

Then, the projection amplitude is given by:
\begin{equation}
    \psi_{A/B, L, l}^{\boldsymbol r_0}(\boldsymbol r) = e^{i  \boldsymbol K_L \boldsymbol r} \sum\limits_{j} \theta^{L, l}_{A/B}(j) \phi^{\boldsymbol r_0 + (-1)^{A/B} \boldsymbol \delta_j}(\boldsymbol r),
\end{equation}
where $\boldsymbol K_L$ is the $K$--point of the layer $L$ (note that the two layers' $K$--points are shifted with respect to each other due to the relative $\theta$--rotation of the layers), $\boldsymbol \delta_j$ are the three vectors connecting an $AB/BA$ stacking region with the three $AA$ neighbors, and the phases $\theta^{L, l}_{A/B}(j)$ can be read off from Fig.\,3 of Ref.\,\cite{PhysRevX.8.031087}. The factor $(-1)^{A/B}$ is $\pm$ for the $A/B$ sublattices. These phases are given for the $\tau = +$ valley, while the $\tau = -$ valley Wannier functions are obtained by complex conjugation.

For the $A$ sublattice the phases $\theta^{L, l}_A(j)$ are given by

\begin{gather}
    \nonumber\theta^{1, A}_A = \{1, \omega^2, \omega\},\;
    \theta^{1, B}_A = \{-1, -1, -1\}, \\
    \theta^{2, A}_A = \{1, 1, 1\}, \;
    \theta^{2, B}_A = \{1, \omega, \omega^2\},
\end{gather}
where $\omega = \exp(2 \pi i / 3)$, and, similarly, 
\begin{gather}
    \nonumber\theta^{1, A}_B = \{-1, -1, -1\}, \;
    \theta^{1, B}_B = \{-1, -\omega, -\omega^2\}, \\
    \theta^{2, A}_B = \{-1, -\omega^2, -\omega\},\;
    \theta^{2, B}_B = \{1, 1, 1\}.
\end{gather}

From the outlined procedure we deduce that the STM images are independent of the spin-component of a Néel order. However, the valley-component affects the STM image. Consider, for instance, the Néel order $\sigma^0 \otimes \tau^v$ with $v = 1,\,2$ or $3$. In case of the order parameter condensation, the wave function is a product state of the local ground states (in the valley space) of the terms $(-1)^{A/B} \tau^v$ that alternate with sublattice index, $|\psi\rangle = |+-+-\ldots+-\rangle$, where $|\pm\rangle$ are the local eigenstates with the eigenvalues $\pm 1$. Using this wave function, we compute the contribution to the individual sites on the graphene scale in the $AA/BB$ regions:
\begin{gather}
    \mathcal{I}^{L, l}(\boldsymbol r) = \sum\limits_{n,\boldsymbol r_0} \left|v^{n,\boldsymbol r_0}_{+} \psi_{l(\boldsymbol r_0), L, l}^{\boldsymbol r_0} + v^{n,\boldsymbol r_0}_{-} \left(\psi_{l(\boldsymbol r_0), L, l}^{\boldsymbol r_0}\right)^* \right|^2,
\end{gather}
where $v^{n,\boldsymbol r_0}_{\pm}$ are the $\pm$--valley components of the local ground state of the $(-1)^{A/B} \tau^v$ Néel order, the summation $\boldsymbol r_0$ runs over all sites on the moiré scale and $n$ runs over the orbitals on a given moiré site, and $l(\boldsymbol r_0)$ is the sublattice index of the site $\boldsymbol r_0$. In this work, we primarily focus on one of the layers, i.\,e., consider $L = 0$.

\section{Interaction terms}
\label{appendix:potentials_shape}
We consider a setup where the electron-electron Coulomb interaction is screened by the two metallic gates put at the distance $d$ from both sides of the moiré sheet.

We use the interaction potential $V_{ij}$ of electrons placed at moiré sites $\boldsymbol R_i$, $\boldsymbol R_j$ as
\begin{gather}
    V_{ij} = \int d\boldsymbol r_i d\boldsymbol r_j |\psi_D^{\boldsymbol R_i}(\boldsymbol r_i)|^2 |\psi_D^{\boldsymbol R_j}(\boldsymbol r_j)|^2 V^{d}(\boldsymbol r_i - \boldsymbol r_j),
\end{gather}
where $|\psi_D^{\boldsymbol R}(\boldsymbol r)|^2$ is the amplitude at position $\boldsymbol r_i$ of a wave function centered around $\boldsymbol R$, and $D$ is the blob width. 

The wave function amplitude, as shown in Appendix\,\ref{appendix:STM}, could be written as
\begin{gather}
\label{eq:blobs}
    |\psi_D^{\boldsymbol R}(\boldsymbol r)|^2 = \beta \phi_D^{\boldsymbol{R}}(\boldsymbol{r}) + \frac{1 - \beta}{3}\sum_j \phi_D^{\boldsymbol R + (-1)^{A/B} \boldsymbol \delta_j}(\boldsymbol r), \\
    \phi_D^{\boldsymbol{R}}(\boldsymbol r ) = \frac{1}{\sqrt{2 \pi D^2}} \exp \left( -\frac{|\boldsymbol r - \boldsymbol R|^2}{2 D^2} \right),
\end{gather}
where, following Ref.\,\cite{PhysRevX.8.031087}, we added an additional central wave-function density $\beta \phi_D^{\boldsymbol{R}}(\boldsymbol{r})$ controlled by the small parameter $\beta$.

The interaction potential is given by
\begin{gather}
    V^d(\boldsymbol r) = \frac{e^2}{\epsilon} \sum\limits_{n=-\infty}^{+\infty} \frac{(-1)^n}{\sqrt{r^2 + d^2 n^2}},
\end{gather}
where $\epsilon$ is the dielectric permittivity. 

We note that at $d \ll D$, electrons can only interact on-site. Therefore, the interaction matrix $V_{ij}$ is given by solely the number of overlapping blobs, giving rise to the well-known $V_0/V_1/V_2/V_3 = 3:2:1:1$ ratio which can be written in terms of the cluster charge interaction Eq.\,\eqref{eq:hamiltonian}. As $d$ grows, the $3:2:1:1$ ratio changes. In Eq.\,\eqref{eq:hamiltonian}, we define $U_0 = V_0 - 3 V_3$ to indicate the deviation from the $3:1$ ratio. To produce the Fig.\,\ref{fig:U0UQ} in the main text, we employed $D = 0.2 L_{\text{M}}$ and $\beta = 0.15$.

\bibliography{refs}
\end{document}